\documentclass[aps,pra,amsmath,amssymb,twocolumn,longbibliography]{revtex4-2}
\pdfoutput=1 
\usepackage{graphicx}
\usepackage{adjustbox}
\usepackage{bm}
\usepackage{color}
\usepackage{braket}
\usepackage{standalone}
\usepackage{multirow}
\usepackage{mathrsfs}
\usepackage{dsfont}
\usepackage[colorlinks,bookmarks=true,citecolor=blue,linkcolor=red,urlcolor=blue]{hyperref}
\definecolor{dgreen}{rgb}{0.1,0.5,0.1}
\definecolor{red}{rgb}{1,0,0}

\usepackage[caption=false]{subfig}
\usepackage{floatrow}

\newcommand{\new}[1]{{\color{black} #1}}
\newcommand{\old}[1]{{\color{red} #1}}

\renewcommand{\H}{\mathcal{H}}

\begin{document}

\title{Correlation-Induced Sensitivity and Non-Hermitian Skin Effect of Quasiparticles}


\author{Tommaso Micallo} 
\email{tommaso.micallo@tu-dresden.de}
\affiliation{Institute of Theoretical Physics, Technische Universit\"at Dresden and W\"urzburg-Dresden Cluster of Excellence ct.qmat, 01062 Dresden, Germany}
\author{Carl Lehmann} 
\affiliation{Institute of Theoretical Physics, Technische Universit\"at Dresden and W\"urzburg-Dresden Cluster of Excellence ct.qmat, 01062 Dresden, Germany}
\author{Jan Carl Budich} 
\email{jan.budich@tu-dresden.de}
\affiliation{Institute of Theoretical Physics, Technische Universit\"at Dresden and W\"urzburg-Dresden Cluster of Excellence ct.qmat, 01062 Dresden, Germany}
\date{\today}

\begin{abstract}
Non-Hermitian (NH) Hamiltonians have been shown to exhibit unique signatures, including the NH skin effect and an exponential spectral sensitivity with respect to boundary conditions. Here, we investigate as to what extent these remarkable phenomena, recently predicted and observed in a broad range of settings, may also occur in closed correlated fermionic systems that are governed by a Hermitian many-body Hamiltonian. There, an effectively NH quasiparticle description naturally arises in the Green’s function formalism due to inter-particle scattering that represents an  inherent source of dissipation. As a concrete platform we construct an extended interacting Su-Schrieffer-Heeger (SSH) model subject to varying boundary conditions, which we analyze using exact diagonalization and non-equilibrium Green's function methods. That way, we clearly identify the presence of the aforementioned NH phenomena in the quasi-particle properties of this Hermitian model system.

\end{abstract}

\maketitle

\section{Introduction}
Leveraging the concept of effective non-Hermitian (NH) Hamiltonians, a number of intriguing phenomena unique to dissipative systems have been recently experimentally discovered and explained with theory in a wide range of physical settings \cite{Rudner2009, Zeuner2015,Lee2016, Gong2018, Zhou2018, Kunst2018,Imhof2018,Yao2018NHBands, Xiao2020,Ashida2020, Weidemann2020,Yuce2020}. This prominently includes NH topological properties such as the winding of generalized energy eigenvalues in the complex plane \cite{Shen2018, Gong2018} entailing the NH skin effect, i.e.,  the accumulation of a macroscopic number of eigenstates at the boundary  \cite{Yao2018,  Lee2019, Borgnia20,Okuma2020,  Helbig20,Znidraric22}, as well as the anomalous sensitivity of surface zero-modes with respect to boundary conditions \cite{Kunst2018, Xiong2018, Okuma19,  Li2020, Koch2020, Budich2020ntos,  koch2022}. 

In most physical scenarios, NH Hamiltonians are used as a conceptually simple tool to effectively model a system-environment coupling \cite{Fukui1998,Ashida2020}, e.g. as an approximation to a quantum master equation. There, complex energy eigenvalues account for decay into a bath and the amplification from gain in an optically active system, respectively \cite{Ashida2020,Menke2017, koch2022}. However, even in closed quantum many-body systems, quasi-particle excitations may exhibit effective dissipation due to their scattering off other degrees of freedom within the considered system \cite{Pitaevskii1980,Rotter2009, Zyuzin2018,  Michishita20, Okuma21}. The imaginary part of the quasi-particle energies, as described by the self-energy $\Sigma$ entering the Green's function (GF), then models their lifetime, and non-trivial matrix structures of $\Sigma$ have been found to give rise to interesting NH spectral properties such as exceptional points \cite{Berry2004,Yoshida2018, Yoshida2020,Nagai2020, Rausch2021, Michen2021, Lehmann2021}.

 \floatsetup[figure]{style=plain,subcapbesideposition=top} 
\begin{figure}[htp!]	 
  \centering 
  \sidesubfloat[]{\includegraphics[width=0.55\textwidth]{model_with_parameters.pdf} \label{model}} 
  \sidesubfloat[]{\includegraphics[width=0.27\textwidth]{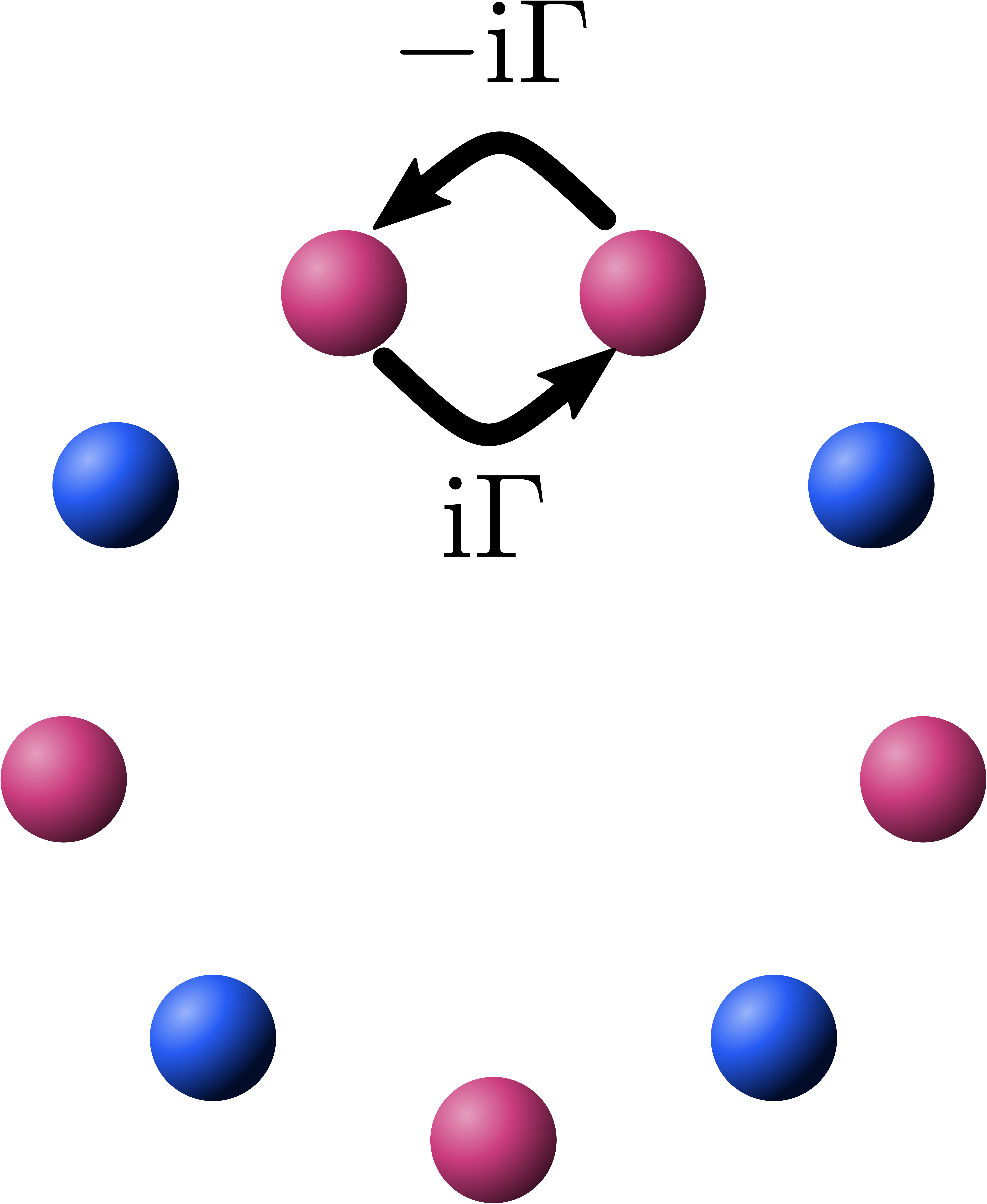} \label{model_gbc} }\\ 
  \vspace{1cm}  
  \sidesubfloat[]{\includegraphics[width=0.9\columnwidth]{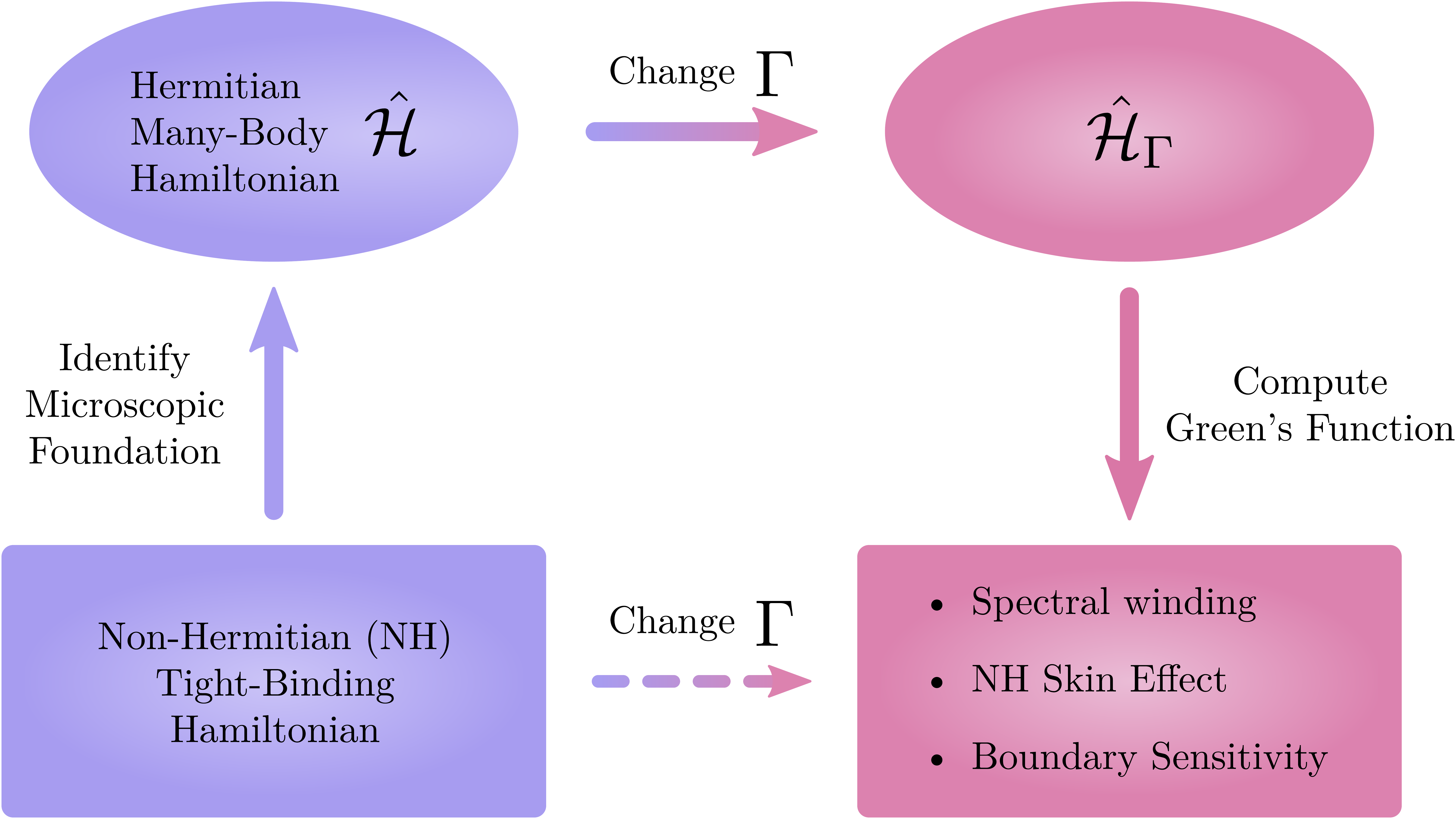} \label{scheme} }
\caption{ \label{fig:model} (a)~Illustration of the tight binding Hamiltonian $\H_0$  (Eq. \protect\eqref{EQ:noninteracting}) with parameters $t_1,t_2$ and two-body interaction $\mathcal{V}$ (Eq. \protect\eqref{EQ:interaction}) with sublattice-specific interaction strenghts $U_\mathrm{AA},U_\mathrm{BB}$ in a two-leg ladder geometry, with sublattice A (B) as the red upper (blue lower) leg. (b)~ Illustration of the broken ring geometry with generalized boundary condition paramter $\Gamma \in [0,1]$, where $\Gamma =0$ ($\Gamma = 1$) corresponds to open (periodic) boundary conditions. (c) Schematic illustration of our present program, where the path of solid arrows is followed to demonstrate its (qualitative) commutation with the beaten dashed path.}
\end{figure}

In this work, we study the NH boundary sensitivity of quasi-particles in a closed quantum many-body system, demonstrating the occurrence of both a NH skin effect and the striking sensitivity of a topological edge-mode with respect to small changes in the boundary conditions, despite the Hermitian nature of the governing many-body Hamiltonian. To this end, we construct and solve a model for a one-dimensional chain of correlated fermions with sub-lattice dependent interactions and varying boundary conditions (see Fig.~\ref{fig:model}(a-b) for an illustration) at finite temperature. In the framework of full exact diagonalization (ED), we extract the NH quasi-particle Hamiltonian directly from the K{\"a}ll{\'e}n-Lehmann representation of the retarded GF, and analyze its NH skin effect. In addition, to access bigger system sizes, we compute the GF within the conserving second Born approximation (SBA), thus finding numerical evidence for a sensitivity  of the lowest lying quasi-particle mode with respect to boundary conditions that scales exponentially in system size. Our results corroborate the picture that quasi-particles behave qualitatively similarly to excitations of an effective NH tight-binding model, even when it comes to changes in the boundary conditions of the underlying many-body Hamiltonian (cf.~commuting diagram in Fig.~\ref{fig:model}(c)), which implies genuinely NH spectral and spatial distribution properties  \cite{Bergholtz2021,Michen2021,Crippa2021,koch2022}.

The remainder of this article is structured as follows: The Green's function formalism underlying the definition of an effective NH quasi-particle Hamiltonian is discussed in Section \ref{sec:GF_formalism}, together with a brief synopsis of the numerical methods used to compute the GF. In Section \ref{sec:model_design}, a microscopic model is constructed so as to mimic a topological NH tight-binding model at the quasi-particle level. This model is analyzed in Section \ref{sec:boundary_results}, demonstrating how the effective NH Hamiltonian responds to varying boundary conditions of the many-body Hamiltonian. A concluding discussion is presented in Section \ref{sec:Conclusions}. In this work, we use the natural units $\hbar=k_\mathrm{B}=1$.

\section{Methods and model building}

\subsection{From Green's functions to Non-Hermitian Hamiltonians \label{sec:GF_formalism}}
In correlated quantum many-body systems, many important physical aspects such as response properties and elementary excitations are encoded in correlation functions \cite{Zagoskin2014}. In particular, equilibrium quasi-particle properties including  the spectral functions and key susceptibilities are described by the single-particle (or two-point) retarded Green's function (rGF). In this framework, the notion of an independent particle described by a non-interacting Hermitian Hamiltonian may be replaced by a quasi-particle governed by and effective NH Hamiltonian $\H_\mathrm{eff}$\cite{Pitaevskii1980,Kozii2017,Rausch2021, Okuma21, Michen2021,Crippa2021}. There, the NH character reflects the finite lifetime of quasi-particles due to inter-particle scattering \cite{Kozii2017}. In the real-space and time domain, the rGF is defined as

\begin{equation}
\label{EQ:green_realtime}
    G^\mathrm{R}_{m,n}(t)=-\mathrm{i}\theta(t)\left< \{ c_m(0), c^\dagger_n (t) \} \right> \mathrm{,}
\end{equation}
where $c(t)$, ($c^\dagger(t')$) is the annihilation (creation) operator in the Heisenberg picture, $m,n$ are indices referring to both spatial and internal (e.g. sublattice) degrees of freedom, and $\left< ... \right>$ is the thermal average\new{ in the grand canonical ensemble} with respect to the full many-body Hamiltonian $\H$. 
Fourier transforming to the frequency domain with respect to $t$, the frequency-dependant effective NH Hamiltonian $\H_\mathrm{eff}(\omega)$ and self energy $\Sigma(\omega)$ are introduced as
\begin{equation}
\label{EQ:green_frequency}
    G^\mathrm{R}(\omega)=\lim_{\eta \rightarrow 0^+}(\omega +\mathrm{i} \eta - \H_\mathrm{eff}(\omega))^{-1} \mathrm{~;}
\end{equation}
\begin{equation}
   \H_\mathrm{eff}(\omega)=\H_0+\Sigma(\omega) \mathrm{,}
   \label{eqn:heff}
\end{equation}
where matrix indices have been dropped for brevity. We focus on quasi-particle excitations close to the Fermi energy ($\omega=0$) and their NH topological properties. To obtain a frequency independent effective NH Hamiltonian, we approximate the self energy by its value $\Sigma=\Sigma(0)$ at the Fermi energy. Generally speaking, since the poles of the rGF (see Eq.~(\ref{EQ:green_frequency})) lie in the lower half of the complex plane, the eigenvalues of $\H_\mathrm{eff}$ can only have nonpositive imaginary parts, corresponding to negative inverse of lifetimes. We now outline how the rGF is practically computed in our present analysis.

\paragraph*{Exact Diagonalization.} For finite chains of modest size, the many-body Hamiltonian $\H = \H_0 + \mathcal V$ with non-interacting part $\H_0$ and interaction potential $\mathcal V$ can be fully diagonalized by solving $\H \left| E_\mu \right>= E_\mu \left| E_\mu \right>$. \new{The diagonalization algorithms used in our simulations are provided by the python library NumPy \cite{Harris2020}. }Then, the rGF can be computed within the K{\"a}ll{\'e}n-Lehmann representation\cite{Wang2012}\new{ in the grand canonical ensemble} as

\begin{equation}
\label{EQ:Lehmann_repr}
\begin{split}
   G_{m,n}(\omega)\underset{\eta \rightarrow 0^+}{=} &\sum_{\mu,\nu}\frac{\mathrm{e}^{-E_\mu/T}}{Z}\\
   &\times \Bigg[\frac{\left<E_\mu |c_m| E_\nu \right>\left<E_\nu |c_n^\dagger| E_\mu \right>}{\omega +\mathrm{i} \eta+ E_\mu - E_\nu}\\
   &+ \frac{\left<E_\nu |c_m| E_\mu \right>\left<E_\mu |c_n^\dagger| E_\nu \right>}{\omega +\mathrm{i} \eta+ E_\nu - E_\mu} \Bigg] \mathrm{ ,}
\end{split}
\end{equation}
\new{where $Z=\sum_\mu \mathrm{e}^{-E_\mu/T}$ is the partition function, and the chemical potential has been set to 0 so that the favored configurations are the ones closest to half-filling.}
Numerically, the limit $\eta \rightarrow 0^+$ of the regularization parameter must be approximated, thus setting a limit to the resolution in frequency (or energy). While conferring to the finite size rGF a characteristic spiky spectral structure, the choice of a small but finite regularization parameter only affects quantitatively, but not qualitatively the results presented in this work. To retrieve the effective Hamiltonian~\eqref{eqn:heff},  the limit $\eta \rightarrow 0$ is withheld, and effectively frequency is treated as a complex parameter with a small but finite imaginary part. Where ED results are shown, we use $\eta=0.015$.

\paragraph*{Second Born approximation.} To access longer chains we use the non-equlibrium Green's function approach in conserving second Born approximation (SBA) \cite{Stefanucci2013,Balzer2013,Aoki2014,Schluenzen2016,Schueler2020}, where the limiting factor is that only chains with weak to moderate interaction strengths can be solved to satisfactory accuracy \cite{Eckstein2011,Schlnzen2016,Schueler2020, Murakami2020}. In this framework, we numerically compute the equilibrium rGF in real time (see Eq.~\eqref{EQ:green_realtime} and Appendix~\ref{sec:appendix_2BA}) by solving the Kadanoff-Baym equations of motion, building on the software package NESSi \cite{Schueler2020}. For sufficiently long times, the amplitude of the rGF is damped by the interaction induced dissipation, thus allowing us to converge the Fourier transform to frequency space in order to extract the effective NH Hamiltonian~\eqref{eqn:heff}. 

\subsection{Microscopic lattice model \label{sec:model_design}}
We now construct a microscopic model for a Hermitian many-body Hamiltonian $\H$ that yields a topologically nontrivial quasi-particle spectrum at the effective NH Hamiltonian level (see Eq.~\eqref{eqn:heff}). In particular, we find it interesting to investigate whether quasiparticles can exhibit a similarly dramatic response to changes in the boundary conditions of $\H$ as excitations in NH topological tight-binding models are known to exhibit.
To this end we \new{ now define the different kinds of boundary conditions used in this work.
\begin{itemize}
\item  Periodic boundary conditions (PBC) are used to model translationally invariant chains with a finite number of sites.
\item Open boundary conditions (OBC) are present if the hopping of the first and last site is suppressed, so the system physically appears as a line with two ends (see Fig. \ref{model_gbc} with $\Gamma=0$).
\item Generalized boundary conditions (GBC) interpolate between OBC and PBC with a parameter $\Gamma \in [0,1]$ and can be though as an imperfect ring (see Fig. \ref{model_gbc}). For systems with an integer number of unit cells (even number of sites for the bipartite lattice considered below), $\Gamma=1$ amounts to PBC.
\end{itemize}
We then}
proceed in two steps. First, we discuss an effective NH tight-binding model that has the conceptually simplest anti-Hermitian part enabling the aforementioned NH topological features. The Hermitian part of this NH toy model is then adopted as the non-interacting part $\H_0$ of our microscopic model. Second, a two-body scattering term $\mathcal V$ is constructed that may induce a self-energy mimicking the anti-Hermitian part of the toy model. This results in the full Hermitian Hamiltonian $\H= \H_0 + \mathcal V$. 

\paragraph*{Target NH tight binding model.} To set the stage, we briefly discuss a NH generalization of the celebrated Su-Schrieffer-Heeger (SSH) model \cite{Su1979,Lieu2018,Budich2020ntos} as an archetype of a  topological  NH tight-binding model that we aim at mimicking further below with the quasi-particle excitations of a Hermitian many-body model. The Hermitian part of the Hamiltonian is given by
\begin{equation}
\label{EQ:noninteracting}
    \H_0=\sum_j t_1 c^\dagger_j \sigma_x c_j + \sum_j \frac{t_2}{2} c^\dagger_j (\sigma_x + \mathrm{i}\sigma_z) c_{j+1} + \mathrm{H.c. ,}
\end{equation}
where $c_j=(c_{j, \mathrm{A}}, c_{j, \mathrm{B}})^T$ is the annihilation operator in unit cell $j$, which has a two-spinor structure in A-B-sublattice space, where the standard Pauli matrices $\sigma_{x}, \sigma_z$ act.

The anti-Hermitian part is assumed to be of the form
\begin{equation}
\label{EQ:target_NH}
    \H_\mathrm{AH}= -\mathrm{i}\sum_j   c^\dagger_j  \left( \gamma_z \sigma_z+ \gamma_0 \sigma_0 \right) c_j \mathrm{ ,}
\end{equation}
where $\gamma_0,\gamma_z \ge 0$ and the causality constraints of the rGF amount to $\gamma_0 \ge \gamma_z$.  Even though the $\gamma_0$-term is a simple vertical shift of the eigenenergies in the complex plane, it is thus of fundamental importance for the physical interpretation in terms of quasi-particles with a finite life-time.
For $\gamma_z \neq 0$, the system exhibits topologically non-trivial properties including a spectral winding around the base point $E_\mathrm{b} = -i \gamma_0$ for periodic boundary conditions (PBC)\cite{Gong2018,Lieu2018}, the NH skin effect for open boundary conditions (OBC) \cite{Bergholtz2021}, and an exponential sensitivity of a spectrally isolated mode with respect to boundary conditions \cite{Budich2020ntos}. We will discuss these intriguing phenomena in more detail when analyzing their occurrence one by one in the quasi-particle spectrum of our microscopic model in Sec.~\ref{sec:boundary_results}. 

\paragraph*{Sublattice dependent interaction.}
From Eq.~\eqref{EQ:target_NH}, we note that the anti-Hermitian part of our targeted NH tight-binding model is diagonal in real space. While $\H_\mathrm{AH}$ is independent of the unit cell (translation invariance), it does depend on sub-lattice for finite $\gamma_z$, i.e. in any topologically non-trivial scenario. Quite intuitively, a contribution of this form to the self-energy may emerge from a sub-lattice dependent interaction potential, which physically corresponds to different scattering rates within particles on A sites as compared to particles on B sites, and thus a sublattice-dependent quasi-particle life-time  \cite{Rausch2021, Lehmann2021}. While momentum-transfer in scattering processes may give rise to a more complicated momentum-dependent profile of the scattering rates, the local self-energy is still expected to have the largest magnitude in a finite temperature system. Guided by this intuition, in the following we solve a microscopic many-body model defined by adding to the free Hermitian Hamiltonian $\H_0 $ (see Eq.~\eqref{EQ:noninteracting}) the sublattice dependent two-body interaction

\begin{equation} \label{EQ:interaction}
    \mathcal{V}=\sum_\alpha\sum_{j}U_{\alpha\alpha} \left( n_{j, \alpha}-\frac{1}{2}\right)\left( n_{j+1, \alpha}-\frac{1}{2}\right)\mathrm{ ,}
\end{equation}
where $\alpha=$A,B and $n_{j, \alpha}= c^\dagger_{j, \alpha} c_{j, \alpha}$ denotes the number operator on sublattice $\alpha$ in unit cell $j$. Without loss of generality, in the following we assume $U_\mathrm{AA} > U_\mathrm{BB}$ to induce a staggered scattering rate corresponding to a finite $\gamma_z$ in Eq.~\eqref{EQ:target_NH}. Our full model described by the Hamiltonian $\H = \H_0 +\mathcal V$ is illustrated in Fig.~\ref{fig:model}(a).
This model preserves the unitary particle-hole symmetry $C$, i.e. $C \H C^{-1} = \H$ which is defined by its action 
\begin{equation}
    C c_{i, \mathrm{A}} C^{-1}=-c^\dagger_{i, \mathrm{A}}, \quad C c_{i, \mathrm{B}} C^{-1}=c^\dagger_{i, \mathrm{B}}
    \label{eqn:phs}
\end{equation}
on the field operators and along with its linearity\cite{Ryu2010}.
Note that this symmetry also imposes constraints on the the quasiparticle spectra via the self-energy, as will be discussed in Sec.~\ref{sec:boundary_results} (cf.~Eq.~(\ref{eq:symheff} - \ref{eq:symspec})).

\section{Numerical results for generalized boundary conditions \label{sec:boundary_results}}
We now discuss our numerical results  for the rGF of the model described by $\H = \H_0 + \mathcal V$ (see Eq.~\eqref{EQ:noninteracting} and Eq.~\eqref{EQ:interaction}) obtained in the framework of full exact diagonalization and conserving second Born approximation, respectively (cf. Sec.~\ref{sec:GF_formalism}). Our goal is to analyze as to what extent intriguing NH topological phenomena hosted by the NH tight binding model $\H_0 + \H_{\mathrm{AH}}$  (see Eq.~\eqref{EQ:noninteracting} and Eq.~\eqref{EQ:target_NH}) can be found in the quasi-particle behavior (as governed by Eq.~\eqref{eqn:heff}) of our microscopic many-body model. Specifically, solving for the rGF at finite temperature $T$ and with generalized boundary conditions $\Gamma \in [0,1]$ (see Fig.~\ref{model_gbc}), we will study the following three related properties: The spectral winding of complex quasi-particle energies (Sec.~\ref{sec:boundary_results_PBC}), the non-Hermitian skin effect (Sec.~{\ref{sec:boundary_results_OBC}}), and the exponentially enhanced sensitivity of a spectrally isolated mode (Sec.~\ref{sec:boundary_results_GBC}).
 
\subsection{Spectral Winding\label{sec:boundary_results_PBC}}
\begin{figure}[t]	 
  \centering 
  \sidesubfloat[]{\includegraphics[width=0.9\columnwidth]{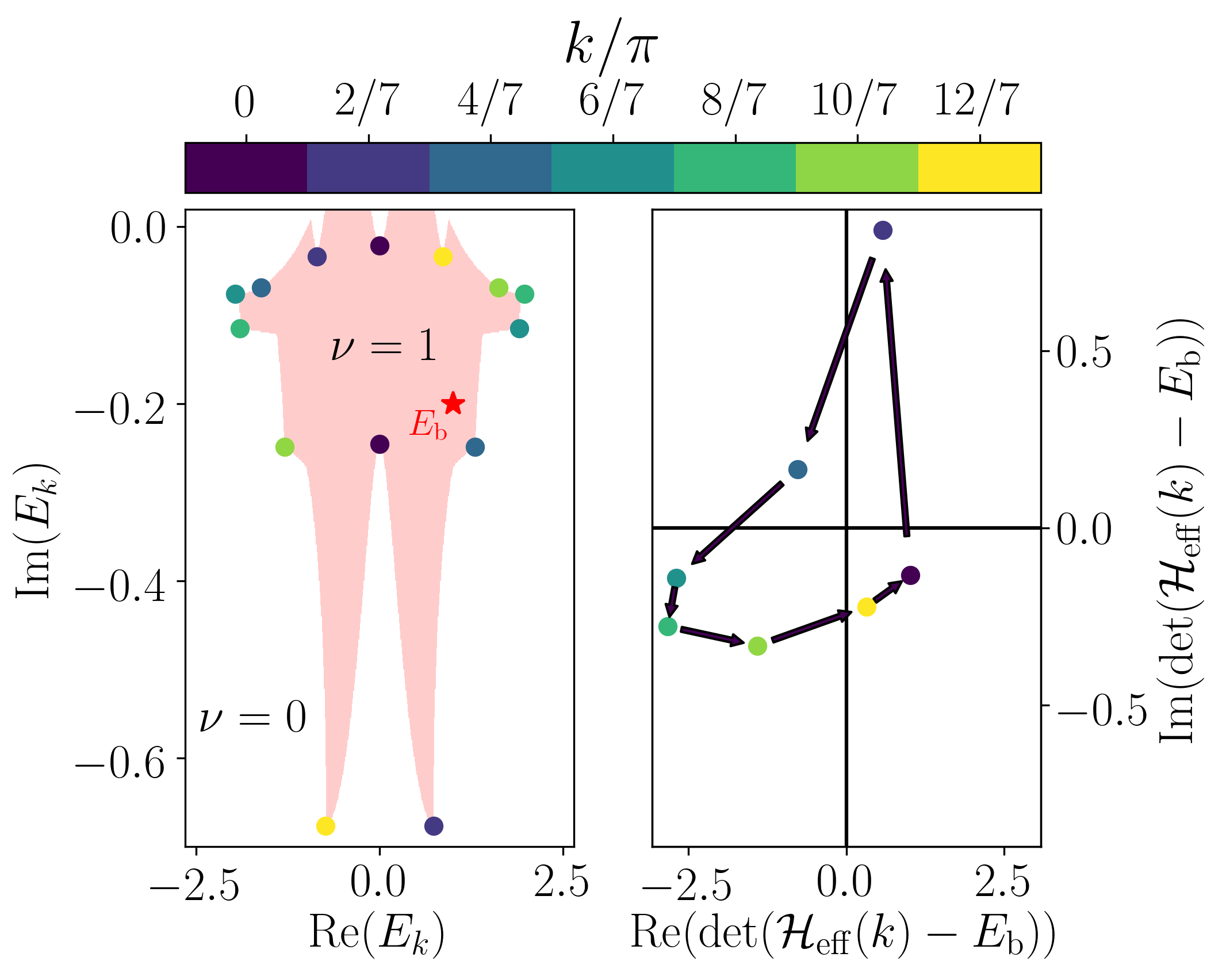} \label{fig:ED_PBC}} \\
  \
  \sidesubfloat[]{\includegraphics[width=0.9\columnwidth]{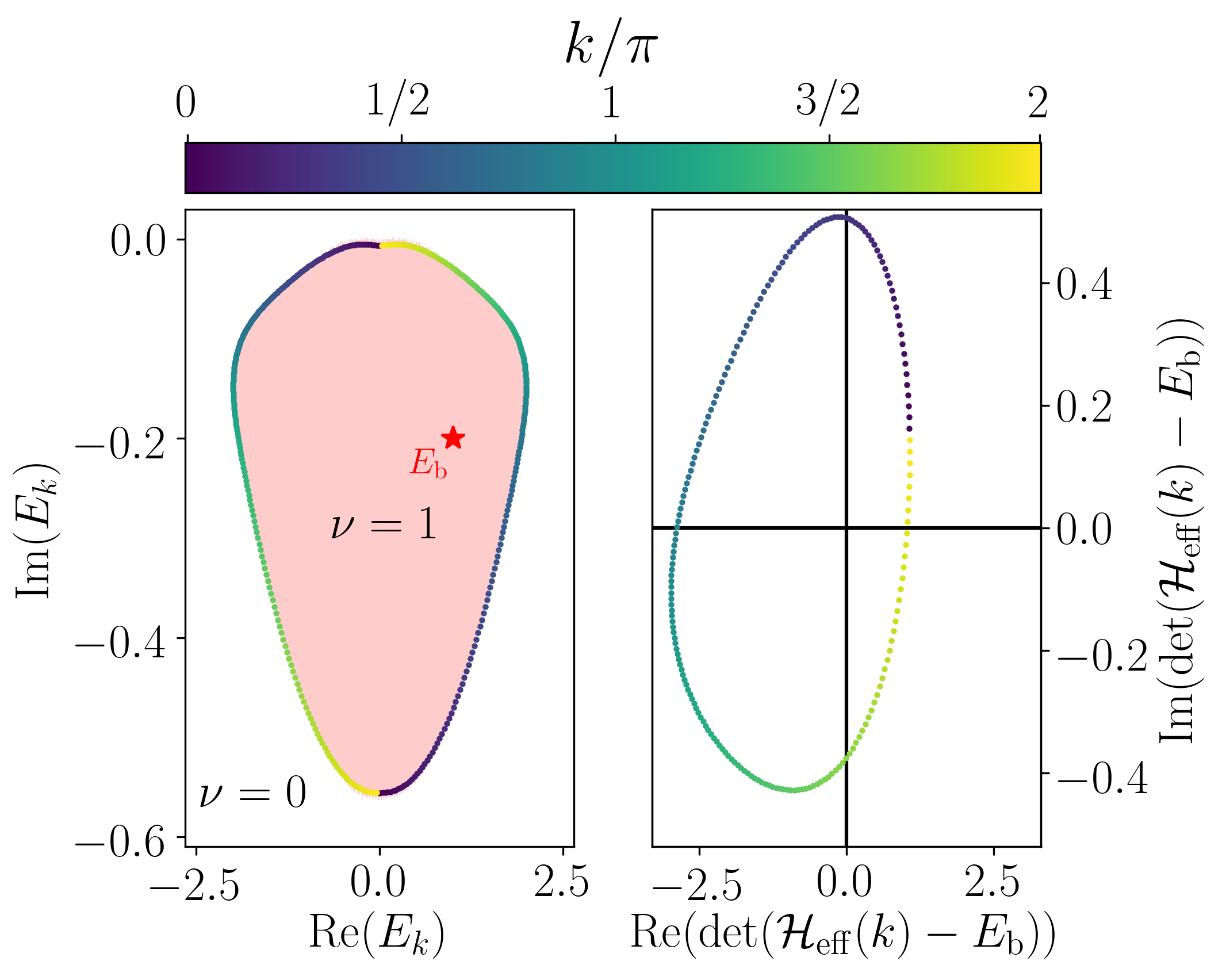} \label{fig:SBA_PBC} }
\caption{ \label{fig:PBC} (a)~Exact diagonalization data ($L=14$ sites) and (b)~Second Born Approximation data ($L=400$ sites) for effective NH Hamiltonians under periodic boundary conditions (PBC). Left panels: complex spectra with PBC showing a point gap, reference point $E_\mathrm{b}= 1-0.2 \mathrm{i}$ denoted as a red star; the red shaded area contains all the reference point, where formally $\nu =1$ holds. Right panels: corresponding nontrivial winding of the determinant around the origin, following Eq.~\eqref{EQ:windingnumber}, with reference point $E_\mathrm{b}= 1-0.2 \mathrm{i}$.  Data obtained for $t_2=-t_1=1$, $U_\mathrm{AA}=1.5$, $U_\mathrm{BB}=0.1$, $T=5$. The arrows and the colors in the right panels show in both cases a counterclockwise winding corresponding to $\nu=1$.}
\end{figure}

We start by considering a translation-invariant system with PBC, such that lattice momentum $k$ is a good quantum number of $\H_\mathrm{eff}$. In this scenario, a topologically non-trivial quasi-particle spectrum is characterized by the winding number \cite{Gong2018}

    \begin{equation}
    \label{EQ:windingnumber}
            \nu  = \oint_{-\pi}^\pi \frac{\mathrm{d}k}{2 \pi \mathrm{i}}\partial_k \log~ \mathrm{det} \left[ \H_\mathrm{eff}(k)- E_{\mathrm{b}}\right]
    \end{equation}

around a fixed base point $E_\mathrm{b}$ in the complex energy plane.
A non-vanishing $\nu$ in system with PBC has been identified as a prerequisite for the NH skin effect \cite{Okuma2020} and the exponentially enhanced boundary sensitivity \cite{Budich2020ntos} to be investigated further below. For finite systems with $L$ unit cells, we compute $\nu$ using a straightforward discretized version of Eq.~\eqref{EQ:windingnumber}, i.e $\nu =  \frac{1}{2 \pi} \sum_n (\mathrm{arg}~ \mathrm{det} \left[ \H_\mathrm{eff}(k_{n+1})- E_\mathrm{b}\right]- \mathrm{arg}~ \mathrm{det} \left[ \H_\mathrm{eff}(k_{n})- E_\mathrm{b}\right] )$ with $k_n = 2\pi n/L;~n=0,1,\ldots , L-1$. We note that $\H_\mathrm{eff}$ inherits the particle-hole symmetry of $\H$ (cf.~Eq.~\eqref{eqn:phs}). Specifically, from the basic definition of the rGF as a correlation function of field operators (see Eq.~\eqref{EQ:green_realtime}), the particle-hole symmetry~\eqref{eqn:phs} can be shown to act on the effective NH $\H_\mathrm{eff}(k)$ as \cite{Chiu2016,Kawabata19, Okuma2022}
    \begin{equation}
        \sigma_z \H_\mathrm{eff}(k) \sigma_z = -\H^*_\mathrm{eff}(-k)
        \label{eq:symheff}
    \end{equation}
    which has the immediate implication 
    \begin{equation}
     \{E_k\} \overset{\mathrm{PH}}{\longleftrightarrow}  \{-E^*_{-k}\}  \mathrm{ .}
     \label{eq:symspec}
    \end{equation}
    on the complex energy eigenvalues, i.e. the imaginary axis is a symmetry axis of the spectrum. This behavior is confirmed by our numerical data (see the left panel in Fig.~\ref{fig:ED_PBC} for ED and Fig.~\ref{fig:SBA_PBC} for SBA data, respectively). These spectra also show a clear point gap that we characterize via the associated winding number. 
\new{A system is considered to possess a point gap if its spectrum does not cross a base point $E_\mathrm{b}$ \cite{Kawabata19,Bergholtz2021}. In particular, spectra that form loops in the complex plane show a point gap with a base point anywhere within the area defined by the loop.}
This NH topological signature highlighted in the corresponding right panels, showing how a phase is accumulated when following Eq.~\eqref{EQ:windingnumber}. While only the larger system sizes treated within SBA render the spectrum a smooth curve in the complex plane, the ED data clearly indicates a qualitatively similar winding behavior already in small systems of $L=14$ sites. From this data, we are able to conclude that, while the self energies obtained with both methods contain additional $k$-dependent terms beyond the simple ansatz in Eq.~\eqref{EQ:target_NH}, the basic intuition of a local self-energy modeled by a finite $\gamma_z$ is sufficient to correctly anticipate the topological winding protected by a point gap.  \new{An example of the full numerically obtained structure of the self-energy is presented in appendix~\ref{sec:additional_structure} and shown in Fig.~\ref{fig:selfenergy_entries} in the case of OBC.}
    
\subsection{Non-Hermitian Skin Effect\label{sec:boundary_results_OBC}}
\begin{figure}[t]	 
  \centering 
  \sidesubfloat[]{\includegraphics[width=0.9\columnwidth]{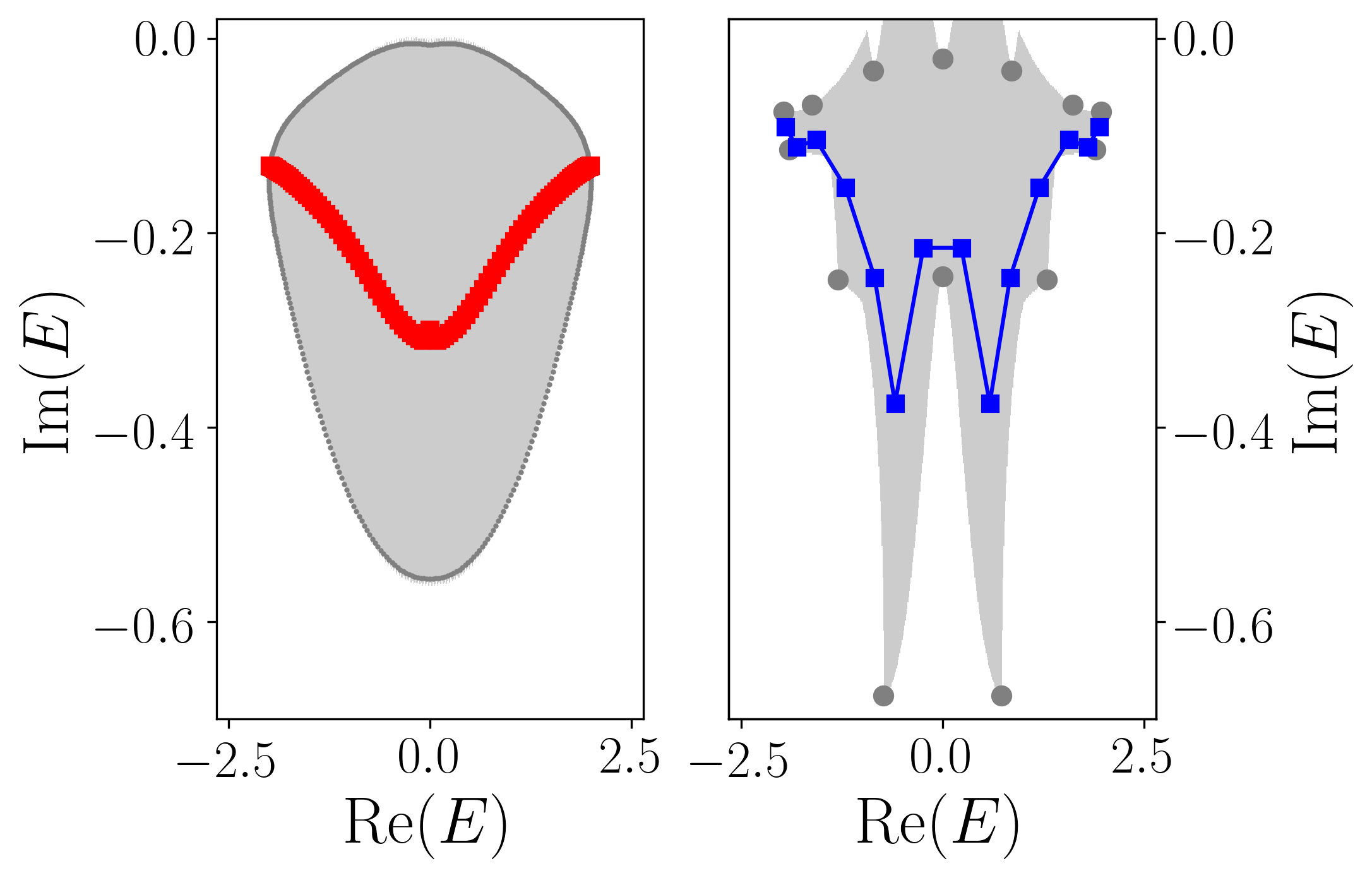} \label{fig:spectra_OBC}} \\
  \
  \sidesubfloat[]{\includegraphics[width=0.9\columnwidth]{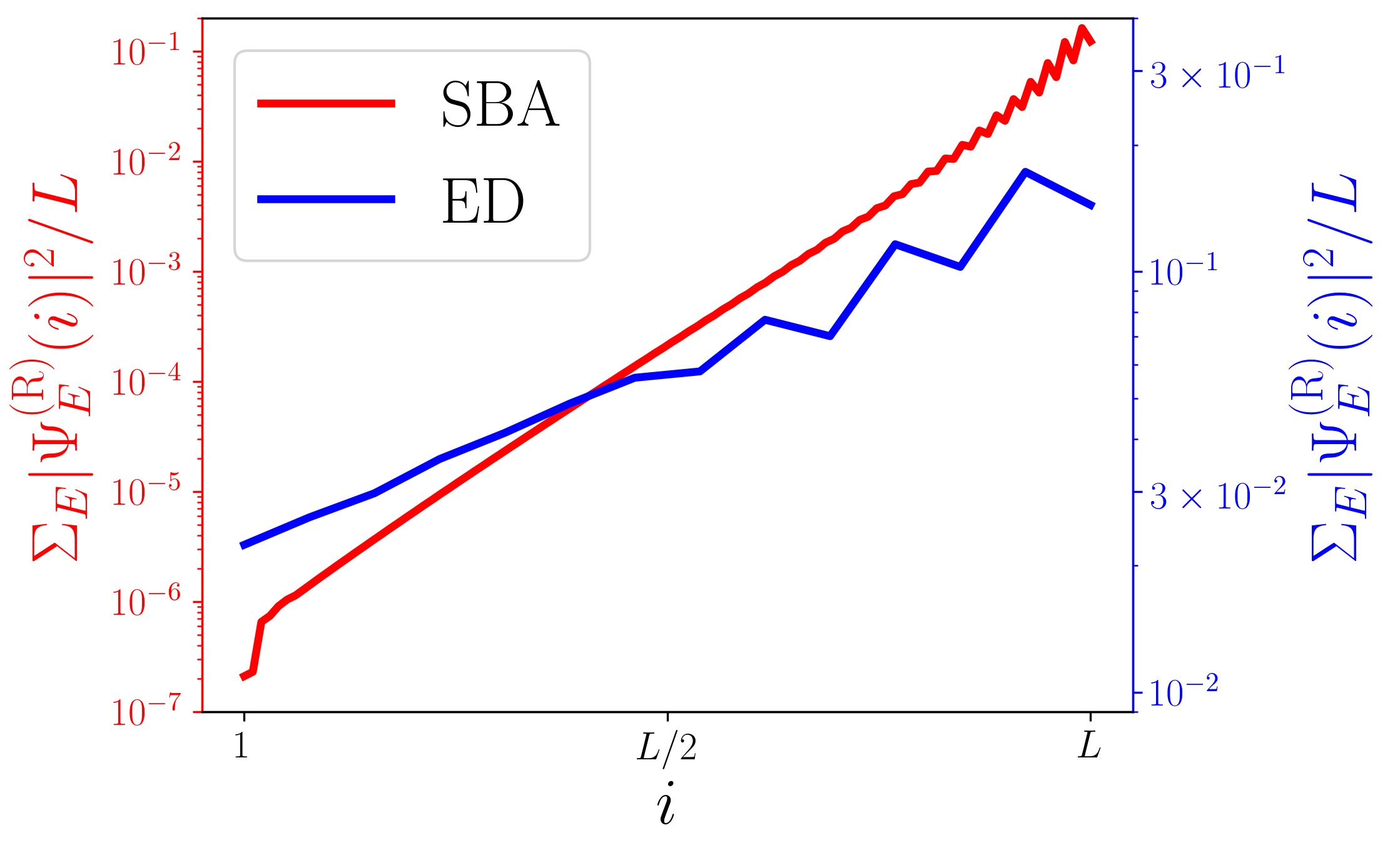} \label{fig:localization} }
\caption{ \label{fig:OBC} (a)~Spectral collapse under open boundary conditions (OBC) for SBA data (red, $L=100$ sites) and ED data (blue, $L=14$ sites), respectively, shown on top of the corresponding PBC data (grey). The spectra drastically change to an arc-shape inside the shaded area associated with $\nu =1$. (b)~Exponential localization of the total spectral weight of right eigenstates as a function of the site index, exemplifying the NH skin effect for SBA data (red, $L=100$ sites) and ED data (blue, $L=14$ sites). Data obtained for $t_2=-t_1=1$, $U_\mathrm{AA}=1.5$, $U_\mathrm{BB}=0.1$, $T=5$.}
\end{figure}

For NH Hamiltonians with $\nu \neq 0$ at PBC, it is well known that the spectrum for OBC may completely change its topology, so as to assume an open arc shape in the complex energy plane that is localized inside the loop defined by the PBC spectrum \cite{Okuma2020, Zhang20}. Moreover, at OBC a macroscopic number of right eigenstates may localize at one end of the chain, while the corresponding left eigenstates will localize at the opposite end \cite{Bergholtz2021}. This is the so-called NH skin effect, the counterpart of $\nu \neq 0$ under OBC \cite{Okuma2020, Zhang20}. While these phenomena have been found and explained with theory in the context of a generalized bulk-boundary correspondence (GBBC) in many NH tight-binding models (see Ref.~\cite{Bergholtz2021} for an overview), here we investigate their occurrence from fully microscopic principles in the effective NH Hamiltonian derived from the rGF of a correlated Hermitian many-body model.  
In other words, in our analysis the boundary conditions are not modified at the level of a NH single-particle Hamiltonian but at the microscopic level of the Hermitian many-body Hamiltonian $\H$. \new{Complementary to our present work, various aspects of the NH skin effect in systems involving interactions have recently been studied~\cite{Kawabata2022,Alsallom2022,Zhang2022,Yoshida2022,Mao2023,Hamanaka2023}}.

In our effective model, using the same parameters and methods as in the previous section on PBC,  in Fig.~\ref{fig:OBC} we observe both the effect of OBC on the complex energy levels collapsing to an arc-shape and demonstrate the occurrence of a NH skin effect for the eigenstates of the effective NH Hamiltonian. Even though this behavior is compatible with intuition about the GBBC occurring in NH tight-binding models, we find it remarkable how the effect of OBC applied to the Hermitian many-body Hamiltonian $\mathcal H$ indeed manifests in the quasi-particle properties according to the principles of NH topology.  Again, the computed OBC self-energies quantitatively deviate from the simple toy model in Eq.~\eqref{EQ:target_NH} so as to include non-local terms \new{(see Fig.~\ref{fig:selfenergy_entries} for an example)}, but the presence of these additional perturbations is not capable of changing the qualitative topological change of the spectrum. The skin effect and the stark change in the eigenvalues when the boundary conditions are varied are key ingredients for the exponential sensitivity analyzed in the following subsection. We stress that also for OBC, the particle hole symmetry of $\H$ imposes the symmetry $\{E\} \overset{\mathrm{PH}}{\longleftrightarrow} \{-E^*\}$ on the complex spectrum of $\H_\mathrm{eff}$.
    
    \subsection{Exponential Sensitivity\label{sec:boundary_results_GBC}}
    
We now turn to considering our model at generalized boundary conditions (GBC) that are used to bridge continuously between open ($\Gamma = 0$) and closed ($\Gamma = 1$) boundary conditions. For systems exhibiting a non-zero spectral winding number $\nu$, \new{the shift in complex energy due to the activation of an end-to-end hopping $\Gamma$ of an isolated mode within the point gap has been predicted to scale exponentially with system size} \cite{Budich2020ntos}. Given that the quasi-particle spectrum of our model system exhibits both $\nu \ne 0$ and the NH skin effect, we find it interesting to study whether even this exponential sensitivity with respect to small changes of $\Gamma$ can be identified in our present context.
    
    \begin{figure}[t]
        \includegraphics[width=0.95\columnwidth]{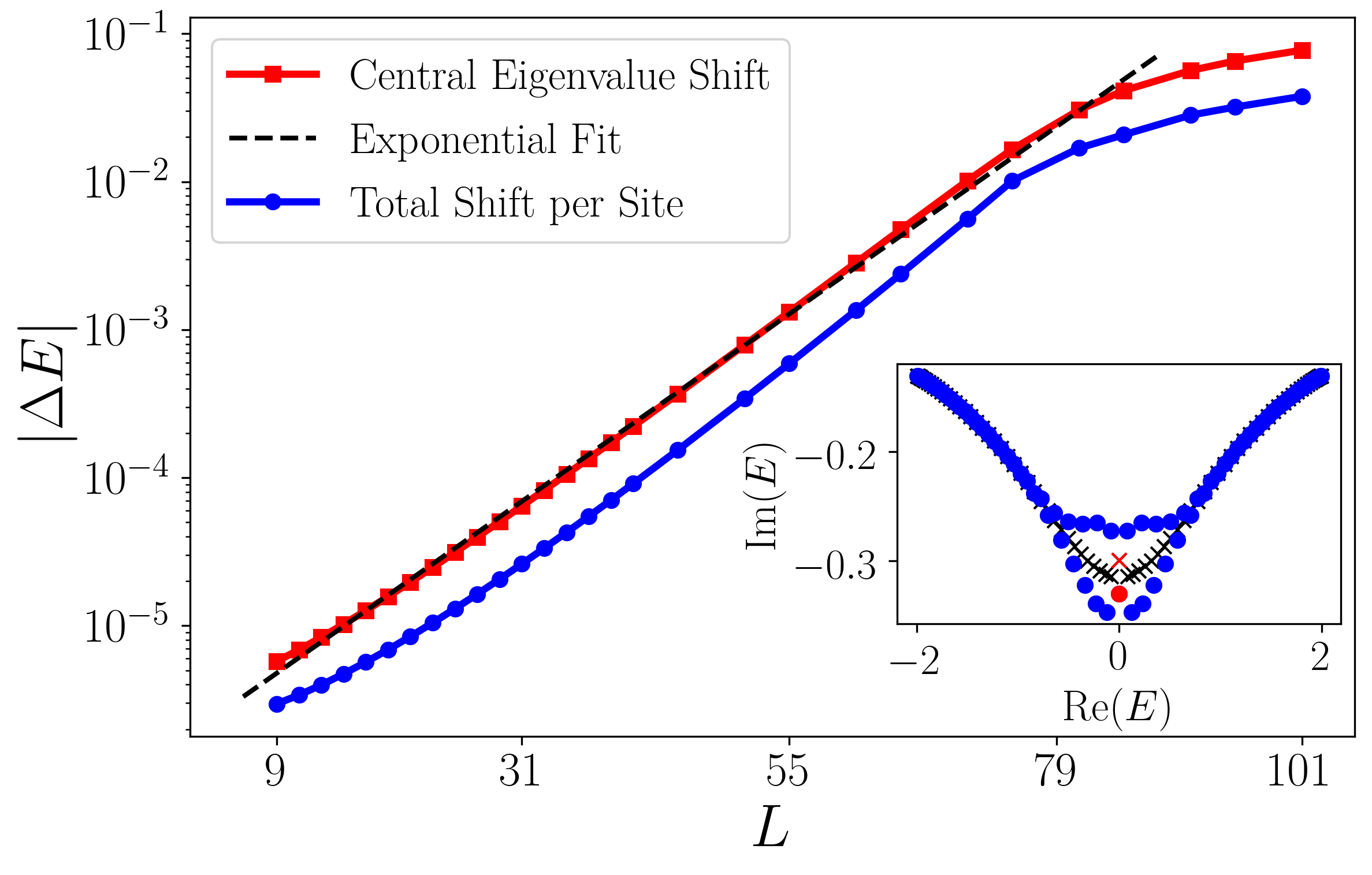}        
          \caption{ \label{fig:exp_sens} \new{Energy shift $\Delta E$ as a function of the number of sites $L$ in response to switching on a small end-to-end hopping $\Gamma=10^{-5}$. The red points show the shift $|\Delta E_0|$ of the spectrally isolated in-gap mode (see Eq.~\eqref{eq:expsen}), and an exponential fit (black dashed line) closely matching our numerical data is plotted as a guide to the eye. The fit excludes the last three point at $L=91,95,101$, where the exponential behavior crosses over to a saturated regime. The blue line shows the overall energy shift $|\Delta E_{\mathrm{tot}}|$ of the spectrum per unit length (see Eq.~\eqref{eq:total_shift}). The inset shows the spectra of the effective Hamiltonian $\H_\mathrm{eff}(\Gamma)$ for $\Gamma=0$ (crosses) and $\Gamma=10^{-5}$ (circles) for $L=81$ sites, where the in-gap mode is highlighted in red. Data obtained via SBA for $t_2=-t_1=1$, $U_\mathrm{AA}=1.5$, $U_\mathrm{BB}=0.1$, $T=5$.}}
    \end{figure}
    
    As shown in Fig.~\ref{model_gbc}, the system under consideration here has an odd number of sites. Since the Hermitian end-connecting term 
    \begin{equation}
        \H_{0,\mathrm{ends}}=\mathrm{i}\Gamma c^\dagger_{N,\mathrm{A}} c_{0,\mathrm{A}} + \mathrm{h.c.}
    \end{equation}
($\Gamma \in \mathbb{R}$) preserves the PH symmetry, the spectrum of $\H_\mathrm{eff}$ is symmetric around the imaginary axis, implying that at least one eigenvalue of $\H_\mathrm{eff}$ must be purely imaginary, independently of the system size. No symmetry-preserving perturbation or interaction can unpin this mode from the imaginary axis, similarly to how a chiral-protected zero mode must always exist in a Hermitian SSH mode with an odd number of sites. 
    \old{This}\new{For this reason, we consider this} generalized imaginary zero-mode \new{to be spectrally isolated; it }solves the eigenproblem $\H_\mathrm{eff}(\Gamma) \left| 0 \right>= -\mathrm{i} E_\Gamma  \left| 0 \right>$ with $E_\Gamma>0$. Adapting the analysis in Ref.~\cite{Budich2020ntos} to this scenario, the spectrally isolated eigenvalue is predicted to shift in response to a change in $\Gamma$ as
    
    \begin{equation}
        \lvert\Delta E\rvert=|E_\Gamma-E_0|=\Delta_0 \mathrm{e}^{\alpha L} \mathrm{,}
        \label{eq:expsen}
    \end{equation}
where $\alpha>0$, i.e. an enhanced sensitivity with increasing system size, is expected for $\nu \ne 0$, while at $\nu =0$ and the absence of the NH skin effect, $\alpha<0$ represents an exponential damping of the level shift \new{$\lvert\Delta E\rvert$}.

\new{It is also possible to introduce a measure of the total spectral response to the activation of the end-to-end hopping $\Gamma$. 
After obtaining the two effective Hamiltonians for zero and non-zero $\Gamma$, one can solve the eigenvalue problems
\begin{equation}
\begin{cases}
\H_\mathrm{eff}(\Gamma) \ket{\psi^\mathrm{R}_j(\Gamma)}=E_{j,\Gamma}  \ket{\psi^\mathrm{R}_j(\Gamma)}\\
\bra{\psi^\mathrm{L}_i(0)}\H_\mathrm{eff}(\Gamma=0) = \bra{\psi^\mathrm{L}_i(0)} E^*_{i,0}.
\end{cases}
\end{equation}
Since the eigenvalues are complex, there is no direct way to infer a correspondence between the sets $\{E_{j,\Gamma} \}$ and $\{E_{i,0} \}$.
We solve this issue by connecting the two sets via maximization of the biorthogonal overlap $\left|\braket{\psi^\mathrm{L}_i(0)|\psi^\mathrm{R}_j(\Gamma)}\right|$\cite{Brody2013,Kunst2018}.
We can then define the total spectral shift
\begin{equation}
\label{eq:total_shift}
 \lvert\Delta E_{\mathrm{tot}}\rvert=\sqrt{\sum_{j=1}^L \frac{(E_{j,\Gamma}-E_{j,0})^2}{L}}.
\end{equation}

}
    
 To clearly identify an exponential scaling, system sizes beyond the scope of full ED are necessary, which is why the data in this section is exclusively based on SBA. Quite remarkably, the effective Hamiltonian computed within SBA clearly exhibits the exponential sensitivity described by Eq.~\eqref{eq:expsen} (see Fig.~\ref{fig:exp_sens}). More specifically, the level shift increases exponentially in systems size over more than three orders of magnitude, before leveling off at a value that is comparable to the size of the point gap of the system along the imaginary axis. This is a strong piece of evidence that topological principles unique to NH systems can manifest in the quasiparticles of many-body systems governed at the microscopic level by a Hermitian Hamiltonian. Similar to the cases on OBC and PBC, the effects on the self-energy of changing $\Gamma$ at GBC are richer than in previously considered NH tight-binding toy models. However, the qualitative similarity of our microscopic results to a simple toy model using the NH term~\eqref{EQ:target_NH} highlights the topological robustness of the studied phenomena.

\section{Conclusions\label{sec:Conclusions}}
In summary, we have designed and studied a Hermitian one-dimensional many-body system of correlated fermions, whose quasiparticle description mimics a topologically non-trivial NH tight-binding model. Specifically, we have investigated several boundary-dependent properties that, at the single-particle level, are unique to NH systems. That way we have demonstrated how changing the boundary conditions of the Hermitian many-body Hamiltonian can indeed have a similar effect on the quasi-particle properties as directly changing the boundary conditions in an effective NH tight-binding model, at least as far as NH topological properties are concerned. In the schematic language of Fig.~\ref{scheme}, we have shown that the dashed and the solid path of the diagram qualitatively commute for proposed model system.

Our findings exemplify on a fully microscopic basis in the realm of quantum many-body physics several intriguing NH topological phenomena that have been predicted at the level of effective NH single-particle models, including the NH skin effect and the exponential sensitivity of spectral properties with respect to small changes in the boundary conditions. These insights suggest that experimental signatures of NH topology can be found in basic spectroscopic experiments and quantum transport settings probing aspects of the single particle Green's function. A closer analysis of the experimental accessibility of the predicted signatures and proposals of platform-specific experimental protocols define interesting subjects of future work.\\

\section*{Acknowledgments}
We would like to thank Lorenzo Crippa, Benjamin Michen, and Alessandro Santini (TM) for discussions. We acknowledge financial support from the German Research Foundation (DFG) through the Collaborative Research Centre SFB 1143 (Project No. 247310070), the Cluster of Excellence ct.qmat (Project No. 390858490), and the DFG Project 419241108. Our numerical calculations were performed on resources at the TU Dresden Center for Information Services and High Performance Computing (ZIH).

\appendix

\section{Second Born Approximation \label{sec:appendix_2BA}}
\begin{figure}[t]
	\centering
	\includegraphics[width=0.7\textwidth]{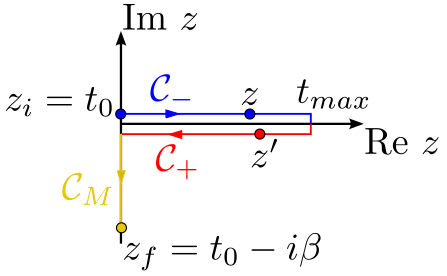}
	\caption{The  Keldysh contour used in our calculations, running from $t_0$ to the maximal time $t_{max}$ back to $t_0$ and ends at the imaginary time $t_0-i\beta$. The complete contour is defined as sum over all time branches: forward $\mathcal{C}_-$, backward $\mathcal{C}_+$ and imaginary(Matsubara) branch $\mathcal{C}_M$. Depending on the time arguments on recovers the ordinary GF like Matsubara or time-ordered component. \label{fig:contour} }
\end{figure}

Using the non-equilibrium Green's functions approach also allows to efficiently compute the rGF in thermal equilibrium.
Working with a generalized GF with complex time arguments $z,z'$ on the Keldysh contour $\mathcal{C}$ (Fig.~\ref{fig:contour}), a perturbative approach in interaction strength using Feynman diagram techniques is taken. Specifically, we the GF is defined as:
\begin{equation}
	\label{eq:selfenergyF}
	\begin{split}
		&G(i,\alpha ,z ;i',\alpha',z') = \\ &-i \frac{\text{Tr}\left(\mathcal{T}_{\mathcal{C}}\exp(-i\int_{\mathcal{C}} dz H(z)) c_{i,\alpha}(z) c_{i',\alpha'}^{\dagger}(z') \right)}{\text{Tr}\left(\exp(-i\int_{\mathcal{C}} dz H(z)) \right)} \text{,}
	\end{split}
\end{equation}
where $\mathcal{T}_{\mathcal{C}}$ denotes the contour time ordering operator, $\int_{\mathcal{C}}$ is the time integral defined on the contour, $H(z)$ is the many-body Hamiltonian which in our case is constant $H(z)=H$ and $c_{i,\alpha}(z)$($c^{\dagger}_{i',\alpha'}(z')$) are the annihilation(creation) operators in the contour Heisenberg picture (a detailed introduction can be found in \cite{Stefanucci2013,Balzer2013,Schueler2020}). 

\begin{figure}[t]
	\centering
	\includegraphics[width=1.0\textwidth]{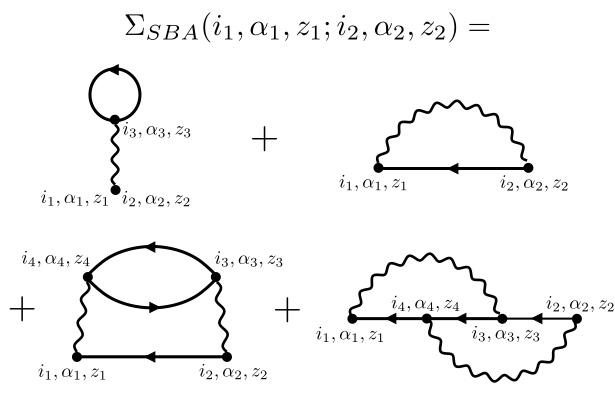}
	\caption{Illustration of the self-energy in the second Born approximation $\Sigma_{SBA}$, which includes all diagrams up to second order (up to two interaction lines). The sum contains the Fock-diagram, Hartree-diagram, exchange-diagram and the direct-diagram (from upper left to lower right).\label{fig:selfenergy_diagram} }
\end{figure}

The GF obeys the Dyson equation, interpreting the site argument $i$ and sublattice index $\alpha$ as a combined matrix index  of $G(z,z^\prime)$, the Dyson equation can be written in a compact form, here in the differential formulation:
\begin{equation}
\begin{split}
	\label{eq:contour_eom}
	\left(i\partial_z -H_0(z) \right)& G(z,z^\prime) = \\
	&\hat{1}\delta_{\mathcal{C}}(z,z^\prime) + \int_{\mathcal{C}}\! d\bar{z}\, \Sigma(z,\bar{z})G(\bar{z},z^\prime) \ ,
\end{split}
\end{equation}
where we have introduced the non-interacting Hamiltonian $H_0(z)$, the delta function on the Keldysh contour $\delta_{\mathcal{C}}(z,z^\prime)$ and the self-energy $\Sigma(z,\bar{z})$.
The self-energy captures all interaction effects and can be written in terms of Feynman diagrams depending on the GF. In our work we consider only diagrams up to second order, which defines the second Born approximation (SBA) $ \Sigma_\mathrm{SBA} = \Sigma_{1\mathrm{F}} + \Sigma_{1\mathrm{H}}+ \Sigma_{2\mathrm{X}} + \Sigma_{2\mathrm{D}}$ (Fig.\ref{fig:selfenergy_diagram}). Note that Fermi lines in the considered diagrams represent full SBA rather than free GFs, as is characteristic for conserving approximations with self-consistently determined GFs. In other words, the SBA contains the so-called Hartree (1H), Fock (1F), exchange (2X), and direct(2D) terms.
Using the interaction in real space $\tilde{v}(i,\alpha;i',\alpha')=(U_A\delta_{\alpha,A} \delta_{\alpha',A} + U_B\delta_{\alpha,B} \delta_{\alpha',B})(\delta_{i,i'+1}+\delta_{i,i'-1})$ each diagram is written as:

\begin{equation}
	\label{eq:selfenergyF}
	\begin{split}
		\Sigma_{1\mathrm{F}}&(i_1,\alpha_1, z_1;i_2,\alpha_2, z_2) = -i \delta_{\mathcal{C}}(z_2,z_1) \delta_{i_1,i_2} \delta_{\alpha_1,\alpha_2} \\
		& \times \sum_{i_3,\alpha_3} \tilde{v}(i_1,\alpha_1;\alpha_3,i_3) G(i_1,\alpha_1,z_3;i_3,\alpha_3,z_3^+),
	\end{split}
\end{equation}

\begin{equation}
	\label{eq:selfenergyH}
	\begin{split}
	\Sigma_{1\mathrm{H}}&(i_1,\alpha_1, z_1;i_2,\alpha_2, z_2) = i \delta_{\mathcal{C}}(z_2,z_1) \\
	&\times  \tilde{v}(i_1,\alpha_1;i_2,\alpha_2)   G(i_1,\alpha_1,z_1;i_2,\alpha_2,z_1^+),
    \end{split}
\end{equation}

\begin{align}
	\label{eq:selfenergyD}
	\begin{split}
	\Sigma_{2\mathrm{D}}&(i_1,\alpha_1, z_1;i_2,\alpha_2, z_2) = G(i_1,\alpha_1,z_1;i_2,\alpha_2,z_2) \\ \times &\sum_{i_3,i_4} \sum_{\alpha_3,\alpha_4} \Big[
	\tilde{v}(i_1,\alpha_1;i_4,\alpha_4) \tilde{v}(i_2,\alpha_2;i_3,\alpha_3)     \\ \times & 
	G(i_3,\alpha_3,z_2;i_4,\alpha_4,z_1)
	G(i_4,\alpha_4,z_1;i_3,\alpha_3,z_2) 
	 \Big],
    \end{split}
\end{align}

\begin{align}
	\label{eq:selfenergyX}
	\begin{split}
	\Sigma_{2\mathrm{X}}&(i_1,\alpha_1, z_1;i_2,\alpha_2, z_2) = \\ -& \sum_{i_3,i_4} \sum_{\alpha_3,\alpha_4} \Big[
	\tilde{v}(i_1,\alpha_1;i_3,\alpha_3)  \tilde{v}(i_2,\alpha_2;i_4,\alpha_4) \\ \times & G(i_3,\alpha_3,z_1;i_2,\alpha_2,z_2)  G(i_4,\alpha_4,z_2;i_3,\alpha_3,z_1) \\ \times & G(i_1,\alpha_1,z_1;i_4,\alpha_4,z_2)  \Big].
    \end{split}
\end{align}
For the first order diagrams (Eq.~\eqref{eq:selfenergyH} and~\eqref{eq:selfenergyF}), the complex time $z^+$ indicates a point of time which happens infinitesimally later than $z$ in the contour time ordering sense.  
Furthermore, the interaction line value $\tilde{v}$ needs to be evaluated carefully with respect to the boundary conditions. 

Our approximation of the self-energy can be derived as a functional derivative of a symmetric functional $\Phi(G)$ ensuring conservation of energy and particle number, if the Dyson equation is solved self-consistently \cite{Stefanucci2013}. 

For practical applications we work in the Keldysh space, which results by employing the Langreth rules and is a decomposition of the complex time GF in several components depending on which branch the time arguments are defined.
Here, using Matsubara, lesser, retarded and left-mixing GFs (see \cite{Stefanucci2013,Schueler2020} for the definition), the Dyson equation becomes a set of computationally challenging coupled differential equations which we solve  numerically using the software package NESSi \cite{Schueler2020}.

\section{Structure of the Self-Energy \label{sec:additional_structure}}
\begin{figure}[t]
        \includegraphics[width=0.95\columnwidth]{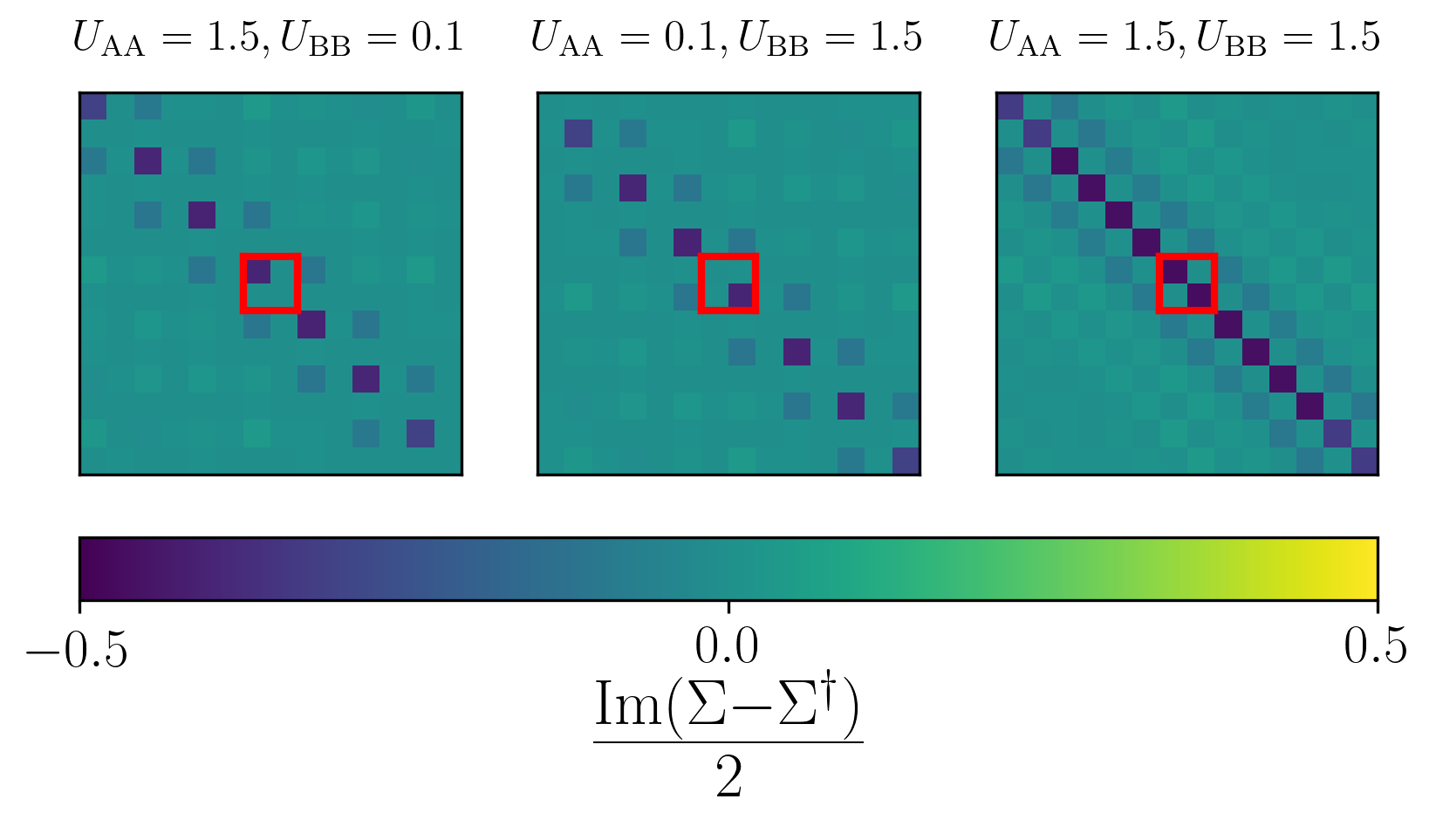}        
          \caption{\label{fig:selfenergy_entries}Entries of the anti-Hermitian imaginary part of the self-energy $\Sigma$ for different values of the interaction parameters $U_\mathrm{AA}$ and $U_\mathrm{BB}$ (indicated above each panel). 
A diagonal block corresponding to a bulk unit cell is highlighted in red to guide the eye.
It is possible to observe how the staggered pattern of interaction (first and second panels) is mirrored in the diagonal entries of the self-energy,  while  the absence of any staggering (third panel) is also inherited by the self energy.
Decomposing the diagonal 2-by-2 blocks on the basis of Pauli matrices one notices that factors of the form of Eq.~\eqref{EQ:target_NH} are by far dominant, with $\gamma_z\simeq 0.1836, -0.1836$ and $\gamma_0\simeq 0.1855, 0.1855$ on average for the first two panels, while for the third one $\gamma_z=0$ within machine error and $\gamma_0=0.4219$.
This figure also shows that the self-energy has a more complex structure than the simple target model of Eq.~\eqref{EQ:target_NH}, since also other (off-diagonal) entries are non-vanishing (though smaller than the dominant diagonal terms).}
    \end{figure}
    
In Section~\ref{sec:model_design} we hinted at how the staggered interaction defined in Eq.~\eqref{EQ:interaction} can give rise to different scattering rates within the different sublattices.
This corresponds to a sizable imaginary $\sigma_z$ term (emergence of finite $\gamma_z$ in the notation of Eq.~\eqref{EQ:target_NH}) on the diagonal blocks of the self-energy matrix, following the same pattern as the staggered interaction.
This behavior is confirmed by the exact numerical data shown in Fig.~\ref{fig:selfenergy_entries}, where we can observe how, whenever $U_\mathrm{AA}\gtrless U_\mathrm{BB}$, i.e. in presence of staggered sublattice-depended interaction, $\gamma_z\gtrless 0$, while if the interactions are equal $\gamma_z$ vanishes.
It is possible decompose the 2-by-2 blocks $h$ on the diagonal of the self-energy on the basis of Pauli matrices, as $h=\sum_i a_i \sigma_i$ via the projection $a_i=\mathrm{Tr}(h \sigma_i)/2$. The $\gamma_{z},\gamma_0$ parameters in Eq.~\eqref{EQ:target_NH} can be extracted as $\gamma_{z(0)}=-\mathrm{Im} a_{z(0)}$. An estimate of the average value of these parameters is also given in Fig.~\ref{fig:selfenergy_entries}.

Fig.~\ref{fig:selfenergy_entries} also shows, as claimed in the main text, how the matrix structure of the self-energy goes beyond the simple terms of Eq.~\eqref{EQ:target_NH}, as those terms appear as diagonal entries, while the interplay between the hopping~\eqref{EQ:noninteracting} and the interaction~\eqref{EQ:interaction} activates also minor off-diagonal scattering channels in the self-energy.


\addcontentsline{toc}{chapter}{Bibliography}

%

\end{document}